\documentclass[pre,aps,twocolumn,showkeys,amssymb]{revtex4}
\usepackage{epsfig}
\usepackage{dsfont} 

\begin{document}

\title{Bidirectional communication in neural networks moderated by a\\ Hebb-like learning rule}
\author{Frank Emmert-Streib}
\email{femmert@physik.uni-bremen.de}
\affiliation{Institut f\"ur Theoretische Physik, Universit\"at Bremen, Otto-Hahn-Allee, 28334 Bremen}

\date{\today}

\begin{abstract}

We demonstrate that our recently introduced stochastic Hebb-like learning rule \cite{vhebb1_2003} is capable of learning the problem of timing in general network topologies generated by an algorithm of Watts and Strogatz \cite{WattsStro_1998}. We compare our results with a learning rule proposed by Bak and Chialvo \cite{cb1999,bc2001} and obtain not only a significantly better convergence behavior but also a dependence of the presentation order of the patterns to be learned by introduction of an additional degree of freedom which allows the neural network to select the next pattern itself whereas the learning rule of Bak and Chialvo stays uneffected. This dependence offers a bidirectional communication between a neuronal and a behavioural level and hence completes the action-perception-cycle which is a characteristics of any living being with a brain.

\end{abstract}

\maketitle

\section{Introduction}

One of the most fascinating complex adaptive systems in nature is the brain. Despite its relatively simple basic units the neurons the cooperative bebaviour of the interconnected neurons and their functional implications are only poorly understood. The problem in investigating this system is not only its complexity, because e.g. the human brain consists of about $10^{12}$ neurons \cite{churchland_1992}, but also its characteristic cycle structure which is known as action-perception-cycle. The difficulty with the action-perception-cycle, which was already known to von Uexk\"ull in 1928 \cite{uexkuell_1928}, is that a closed formulation of the problem has to include a coupled description of the brain and the environment because the actions of an animal are transformed by the environment to perceptions which are transformed by the brain to actions and so on. From this it is also clear that neither the perceptions nor the actions occurring in the system are randomly generated.

In this paper we address the question: How is the learning dynamics of a neural network affected by different mechanisms for the selection of an action? Because learning in neural networks is modulated by a learning rule for the modification of the synaptic weights one can ask more precisely, if the learning rule itself is concerned by the action-selection mechanism.

We approach this problem by comparing two different biologically motivated learning rules for neural networks. The first was proposed by Bak and Chialvo \cite{cb1999,bc2001} and combines experimental findings of Frey and Morris \cite{freymorris_1997} about {\it{synaptic tagging}} with a global reinforcement signal which can be interpreted as a dopamin signal e.g. as in the experiments of Otmakhova and Lisman \cite{otmakhova_1998}. The second was introduced by the author \cite{vphd_2003,vhebb1_2003} and extends the ingredients above by the results of Fitzsimonds \cite{fitzsimonds_1997} about heterosynaptic {\em{long-term depression}} (LTD) which can be qualitatively explained by our stochastic learning rule. Both learning rules are local in the sense that the information, which is used for the synaptic modification, is only provided by the neurons which enclose the synapse and hence can be interpreted as extentions to the classical Hebbian learning rule \cite{h1949}.

As problem to be learned we choose the problem of timing, e.g. catching a ball, in a recurrent network topology which is generated by an algorithm of Watts and Strogatz \cite{WattsStro_1998}. This network class was chosen because the topology is generated in dependence of one parameter, the so called rewiring parameter, and allows to convert a regularly connected network continously in a random one. Recently of special interest was the regime between these two extrema, called small world networks, which could be brought in contact with experimental results about the neuroanatomic structure \cite{latoramarch_2001,latoramarch_2002,WattsStro_1998}.

This paper is organized as follows. In section \ref{model_timing} we define our model. Section \ref{results} demonstrates the practical working mechanism exemplified in learning the problem of timing in a recurrent neural network. We compare the learning behavior of our learning rule \cite{vphd_2003,vhebb1_2003} with the learning rule of Chialvo and Bak \cite{cb1999,bc2001} in dependence of two different action-selection-mechanisms. The paper ends in section \ref{conclusions} with conclusions and an prospect on future work.


\section{The model}\label{model_timing}

If one wants to investigate the learning dynamic of a neural network one has to define every item of table \ref{gen_sys} which characterizes the entire system. 
\begin{table}[h!]
\begin{tabular}{cl}
1.& neuron model\\
2.& topology of the neural network\\
3.& network dynamics\\
4.& learning rule\\
5.& environment\\
6.& interaction of the TBM with the environment\\
\end{tabular}
\caption{\label{gen_sys}Characterization of the entire system}
\end{table}
Metaphorically the points 1. to 4. define the brain of an animal. Because of our simplified description we call this the Toy-Brain-Model (TBM). The concrete definition for each part are as follows. 

1.) Neuron model: Binary neurons $x_i\in\{0,1\}$ with $i\in\{1,\ldots,N\}$. 2.) Topology of the neural network: The neural network is given by the construction algorithm of Watts and Strogatz \cite{WattsStro_1998} with $N=200$ neurons and $k=10$ synapses of each neuron in dependence of a rewiring parameter $p_{rw}$ which regulates the disorder in the network whereas $p_{rw}=0$ corresponds to a regularly and $p_{rw}=1.0$ to a randomly connected network. We choose the construction algorithm of Watts and Strogatz \cite{WattsStro_1998} because real brains are neither regularly nor randomly connected networks but somewhat in between. Experimental results about the neuroanatomical structure concerning cortico-cortical connections in the macaque and cat \cite{scannel_1997,scannel_1995} as well as neuro-neuro copplings in {\em{C. elegans}} \cite{achacoso_1992} indicate that there is a parameter range of the rewiring parameter $p_{rw}$ which is compatible with these experimental findings \cite{latoramarch_2001,latoramarch_2002,WattsStro_1998}.

3.) Network dynamics (winner-take-all): The inner field  of the neurons is calculated by 
\begin{eqnarray}
h_j&=&\sum_i^{\mathrm{all}} w_{ji}x_i  \label{nwta_c4}\\
\end{eqnarray}
Here ``all'' indicates that the summation is carried out over all connected neurons. From the obtained inner fields $h_j$ we select the biggest one 
\begin{eqnarray}
i_\mathrm{max}&=&\mathop{\mathrm{argmax}}_{i}(h_i) \label{nwtas_c4}
\end{eqnarray}
and set the corresponding neuron activity to one and the remaining ones to zero.
\begin{eqnarray}
x_i=\left\{ \begin{array}{r@{\quad}l}
1,  & i=i_{\mathrm{max}}  \\0,  & i\not=i_{\mathrm{max}}
\end{array}\right.
\end{eqnarray}
For this the network dynamics is called winner-take-all mechanism because only the neuron with the highest inner field becomes activated. This kind of network dynamic correpondes in a biological terminus to lateral inhibition.

4.) Learning rule: We choose two different learning rules to adjust the synaptic weights of the neural network and compare them in the result section. 4.a) The learning rule of Chialvo and Bak \cite{cb1999,bc2001} depresses the weights of the active synapses 
\begin{eqnarray}
w_{ij}{\rightarrow}w'_{ij}=w_{ij}-{\delta},
\end{eqnarray}
with $\delta\in[0,1]$, only if the output of the network was wrong indicated by the reinforcement signal $r=-1$ which is democratically fed back to all synapses in the network. The synapses are called active if they were involved in the last signal processing. With the notation introduced by Klemm, Bornholdt and Schuster \cite{kbs2000} this can also be expressed by $\Theta=0$ and $\delta\in[0,1]$ \footnote{It is important to choose $\delta$ from a distribution and not to keep it fixed because this is an additional degree of freedom of the learning rule.} whereas $\Theta$ is a synaptic counter of a certain length which stores the past reinforcement signals.

4.b) Stochastic Hebb-like learning rule: We introduced in \cite{vphd_2003,vhebb1_2003} a novel stochastic learning rule and present here a simplified version which is a special case of \cite{vhebb1_2003} with one degree of freedom less. 

Similar to \cite{cb1999,kbs2000} only active synapses $w_{ij}$ can be updated if $r=-1$ which corresponds to a wrong network output. But now a synapse is updated with the probability $p_\mathrm{\widetilde{c_{ij}}}^\mathrm{rank}$ which is given by \ref{p_rank}. Then the synaptic weights are depressed by
\begin{eqnarray}
\label{synw1}
w_{ij}{\rightarrow}w'_{ij}=w_{ij}-{\delta},
\end{eqnarray}
with $\delta\in[0,1]$.

The stochastic update condition is based on neuron counters $c_{i}$ assigned to all neurons whose dynamics is given by
\begin{eqnarray}
\label{synmem2}
c_{i}{\rightarrow}c`_{i}=\left\{ \begin{array}{r@{\quad}l}
{\Theta}, if &c_{i}-r>{\Theta}   \\c_{i}-r, if & {\Theta}\ge c_{i}-r\ge 0\\ 0, if & 0>c_{i}-r.
\end{array}\right.
\end{eqnarray}
Here ${\Theta}\in\mathds{N}$ is the memory length of the neuron counters and $r=\pm1$ a reinforcement signal. Equation \ref{synmem2} concerns only the active neurons. The other neuron counters remain unchanged.  

The probability $p_\mathrm{\widetilde{c_{ij}}}^\mathrm{rank}$ of the stochastic update condition is obtained by the evaluation of the following procedure:
\begin{enumerate}
\item Calculate the approximated synaptic counters $\widetilde{c_{ij}}$ of the active synapses by the neuron counters \ref{synmem2},
\begin{eqnarray}
\widetilde{c_{ij}}=c_i+c_j
\end{eqnarray}
\item Because of $c_i\in\mathds{N}$ holds for all $i\in\{1,...,N\}$  $\Rightarrow\widetilde{c_{ij}}\in\mathds{N}$ one can calculate for each active synapse an approximated synapse counter and by this one can assign a probability $p_{\widetilde{c_{ij}}}^\mathrm{rank}$, which is given by the rank ordering distribution
\begin{eqnarray}
&P_{k}^\mathrm{rank} & \propto  k^{-\tau}  \label{p_rank}\\
&k  \in &\!\!\!\!\! \{1,\dots  ,2\Theta+3\}\\   
&\tau  \in &\!\!\!\!\! \mathds{R}^+
\end{eqnarray}
with the mapping $k=2\Theta+3-\widetilde{c_{ij}}$, motivated by \cite{bp2001}. 
\end{enumerate}
For the following simulations we used a neuron memory of length $\Theta=3$ and chose the exponent of the rank ordering distribution to $\tau=2.0$.

5.) Environment: The problem to be learned is a mapping from input neurons to output neurons. As input (output) neurons we define $x_{5(m-1)+1}$ ($x_{100+(m-1)5}$) for $m\in M$ patterns. The mapping consists in a connection from input neuron $x_{5(m-1)+1}$ via inter neurons to output neuron $x_{100+(m-1)5}$ in exactly $T_c=4$ time steps. Arriving sooner or later at the predefined output neuron is assumed as wrong network output. For this we call the problem to be learned timing. 

The difficulty of the problem to be learned is the recurrent topology of the network. In contrast to multilayer feedforward neural networks where the output neurons in the last layer are always reached after $\#layer$ time steps this is not the case for recurrent networks. Hence the problem is not only to reach the predefined output neuron but to reach it exactly after a predefined number of time steps.

A similar problem has been studied in a series of papers by \cite{sb1995,cb1999,bc2001}. But in contrast they used a random topology of the neural network and learned a mapping within a predefined time which is easier because, e.g. direct connections from input to output neurons are not forbidden. However, this is for two reasons not desirable. First, the brain of animals is divided in different areas which are specialized to certain performances, e.g. sensor or motor cortex which correspond to our input and output neurons. However, between these parts there is no direct connection. They are connected via inter neurons which are themselves parts of other specialized areas, e.g. the hippocampus for the consolidation of the memory. Second, there are natural problems an animal is faced which can only be solved by correct timing of the animal's motor action. E.g. monkeys have to catch a branch to the right time to prevent them from falling from the tree. 

6.) Interaction of the TBM with the environment: For the interaction of the TBM with the environment we choose two different strategies which are compared in the result section. 6.a) action-selection-mechanism (ASM) I.: The patterns are independently presented with equal probability. 6.b) action-selection mechanism (ASM) II.: In this case the TBM is equipped with an attribute which allows to choose one of the numbered $M$ patterns explicitely. Moreover, this action (pattern) selection mechanism has a memory of length $M$ to store the results of the last $M$ outcomes. Initially pattern 1 is chosen by the action-selection-mechanism of the TBM as long as the mapping is learned which is signed by the reinforcement signal $r=1$. Then the next pattern with number 2 is selected and the procedure is repeated until both mappings are learned. For this we need the memory to store the last outcomes. This procedure goes on until all $M$ patterns are learned. We emphasize that the patterns are always sequentially presented according to their number. A metaphorical visualization of action-selection-mechanism II. can be given as a possible strategy of learning words of a foreign language. If one wants to learn a certain number of words one would not randomly choose but selectively. Which strategy is the best for oneself is individually different but to go on in learning first if one learned some words correctly seems to be very appealing. 

The notations ``environment'' and ``action-selection-mechanism'' were chosen to indicate that we are trying to describe a simple but natural situation in which an animal interacts with its environment to solve some problem which it faces. Hence the interactions with the environment are not random but based on the preceding experience which is accumulated in the brain. So action-selection-mechanism I. looks from a mathematical point of view naturally but it is completely unconditioned from the state of the neural network and subsequently from the information which was gathered. Action-selection-mechanism II. is a first step to connect the neural network with the presentation statistic of the patterns which has to be generated in a self-organized way by the animal itself. The cyclic connection from the perception of a stimulus to the selection of an action is called action-perception-cycle and is a characteristic of all living beings.



\section{Results}\label{results}

In this section we present the results for the model defined in section \ref{model_timing}. More exactly, we want to investigate the learning behavior of the neural network in dependence of the rewiring parameter $p_{rw}$ of the network topology, the action-selection-mechanism and the used learning rule to adapt the synaptic weights. The question that naturally arises now is, how to evaluate the performance of the network? This is no trivial question because there is no explicit costfunction defined in our model which is minimized during the learning process, e.g. in learning by back-propagation \cite{werbos_1974,rumelharthinton1_1986} in artificial neural networks. Instead we use a rule-based adaptation mechanism in form of Hebb-like learning rules 4.a) and 4.b) which are from a biological point of view plausible. To overcome this problem we introduce virtually an outer observer which observes the entire system and hence possesses any information occurring in the system. Mathematically this is done by an identical copy of the entire system in table \ref{gen_sys}. However, with $\delta=0$ which prevents further learning during the evaluation procedure. The patterns can then be presented in an arbitrary order because there are no correlations between them and we determine each time step
\begin{eqnarray}
E(t)_\mathrm{iabs}=\frac{\# of\ patterns\ learned\ up\ to\ time\ step\ t}{M}\label{iabs_error}
\end{eqnarray}
This is the individual absolute error (iabs) of one network at time point t. Individual indicates that this measure is up to now not averaged over an ensemble simulation. Because the synapses of the network are randomly initialized and the synaptic alterations $\delta$ are chosen randomly from $[0,1]$, $E(t)_\mathrm{iabs}$ is a stochastic process for which the first passage time $T_{\mathrm{FPT}}$, when $E(t)_\mathrm{iabs}$ reaches for the first time zero, is a well defined random variable. 

We choose the first passage time $T_{\mathrm{FPT}}$ at the threshold $E_\mathrm{iabs}=0$ and its distribution $p^{\mathrm{FPT}}$ to evaluate the performance of an ensemble of networks. From the distribution $p^{\mathrm{FPT}}$ one can derive two quantitative measures. First, the mean first-passage time $<T_\mathrm{FPT}>$ given by
\begin{eqnarray}
<T_{FPT}>=\sum_{t'=0}^{\infty} t' p_{t'}^{FPT}
\end{eqnarray}
We omit the indices for the value of the threshold because we only investigate the case $E_\mathrm{iabs}=0$. Second, a measure for the speed of the convergence, the distribution function.
\begin{eqnarray}
P^E(t)=\sum_{t'=0}^{t} p_{t'}^{FPT}
\end{eqnarray}
The distribution function $P^E(t)$ is restricted between $0$ and $1$ and indicates the percentage of the networks which did learn the mapping of all patterns up to time point $t$.

\begin{figure}[h!]
\begin{minipage}[c]{0.45\textwidth}
\epsfig{file=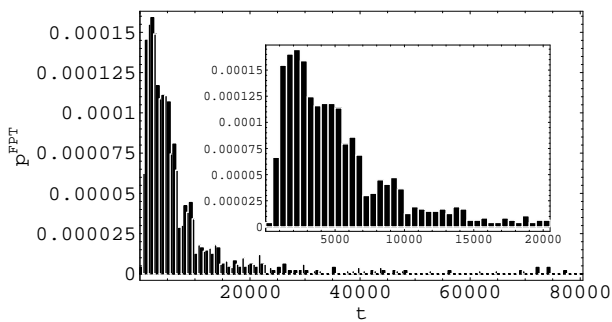, width=1.0\textwidth}
\end{minipage}
\caption{\label{hist_v10_rand3}Distribution $p^\mathrm{FPT}$ of the first-passage times for learning rule 4.b), rewiring parameter $p_\mathrm{rw}=1.0$ and $M=3$ patterns, generated by ASM I. The histogram, with bin width $500$, was generated by simulations over an ensemble of size $N=1000$. The inner figure is a magnification of the first $20000$ time steps.}
\end{figure}

Figure \ref{hist_v10_rand3} shows exemplary the distribution $p^{\mathrm{FPT}}$ of the first-passage times. The shape of $p^{\mathrm{FPT}}$ is characteristic and reflects by a long tail that some networks need much more time to learn the mapping then others.

\subsection{Action-selection-mechanism I.}

\begin{figure}[h!]
\begin{minipage}[c]{0.4\textwidth}
\epsfig{file=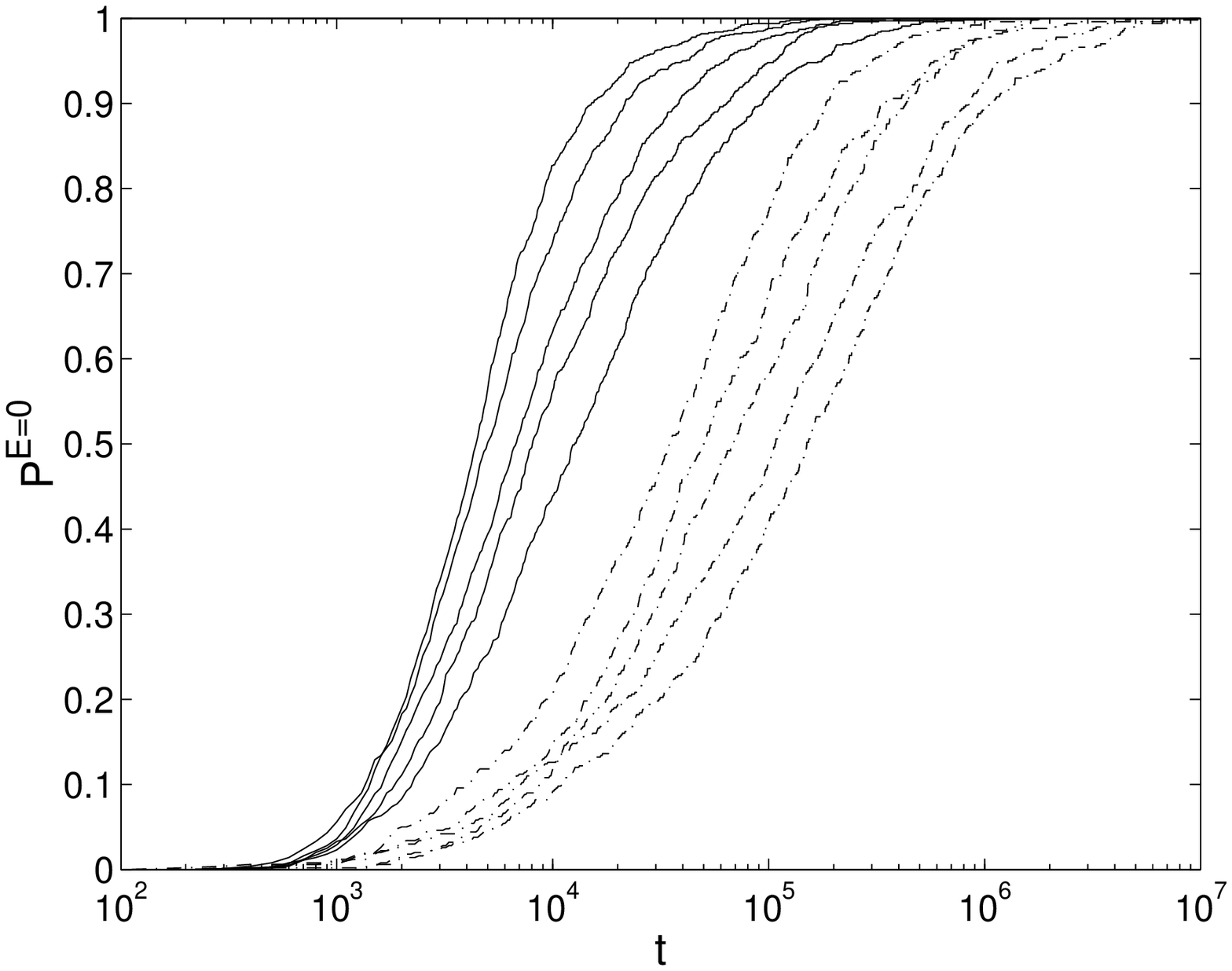, width=1.0\textwidth}
\end{minipage}
\begin{minipage}[c]{0.4\textwidth}
\epsfig{file=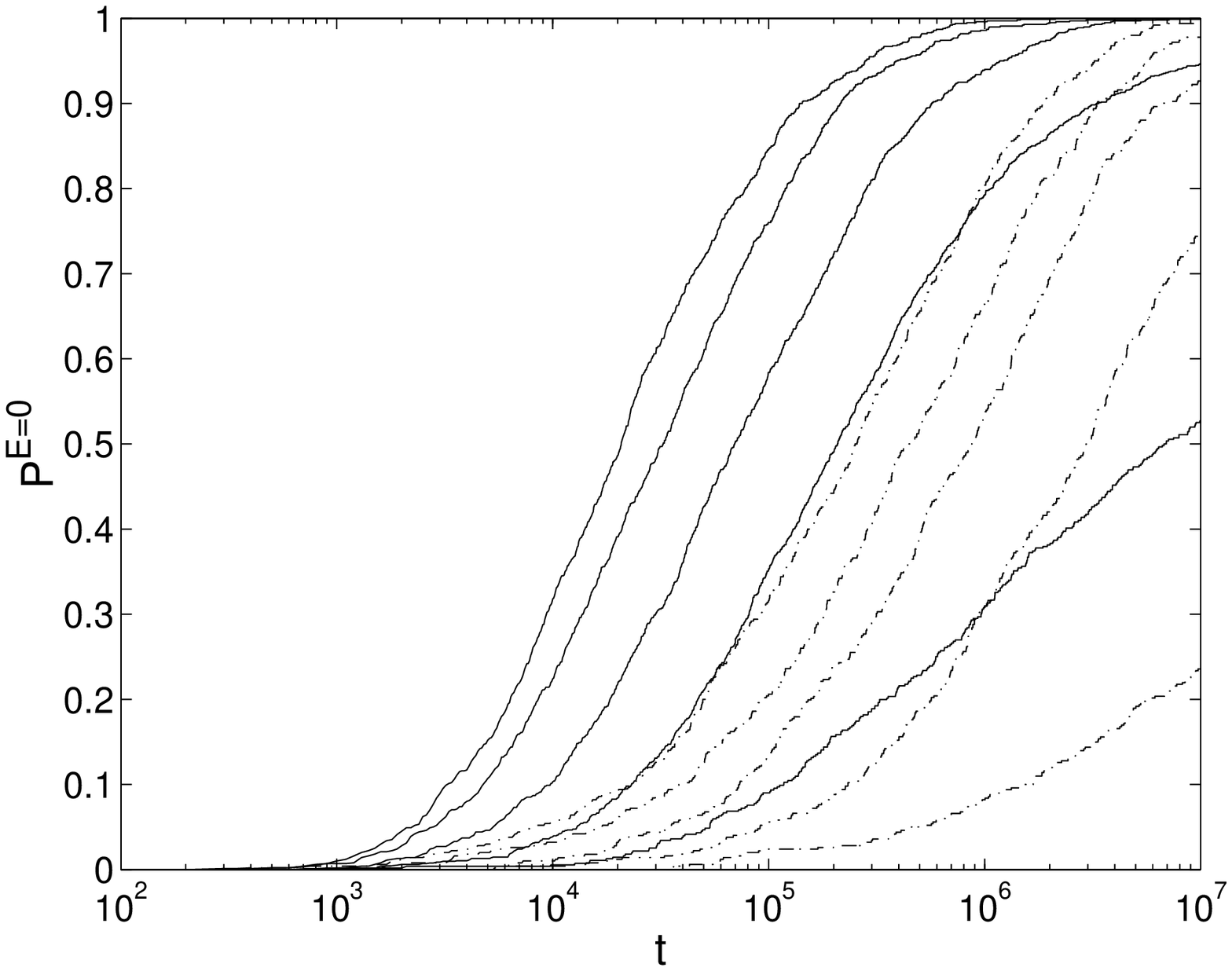, width=1.0\textwidth}
\end{minipage}
\caption{\label{asm1_m3_pl}Distribution function $P^{E=0}(t)$ for $M=3$ and learning rule 4.a) (dash-dot line) and 4.b) (full line) which were obtained for ASM I. The curves are parametriesed from above to below from $p_\mathrm{rw}=1.0$ to $p_\mathrm{rw}=0.6$ (upper figure) and from $p_\mathrm{rw}=0.5$ to $p_\mathrm{rw}=0.1$ (lower figure).}
\end{figure}

{\bf{$M=3$ patterns:}}\\
Figure \ref{asm1_m3_pl} shows the distribution function $P^E(t)$ for learning rule 4.a) (dotted lines) and 4.b) (full lines) in dependence of the rewiring parameter $p_{rw}$. It is clear to recognize that the convergence behavior for learning rule 4.b) is always significantly better. In general holds the less $p_{rw}$ becomes the longer it takes to converge. This is due to the fact that for $p_{rw}=0$ the network is regularly connected without any shortcuts between further remote neurons. Hence there is no path which connects the input with the output neurons within $T_c=4$ time steps. If one increases $p_{rw}$ there exists more and more such shortcuts and the problem can be learned more easily. For $p_{rw}>0$ the problem consists not only in finding connecting paths between input and output neurons but also in preserving paths for already correctly learned mappings. This interplay between path exploration and path conservation makes the problem hard especially for low values of the rewiring parameter.

Table \ref{list_ens1} gives in the second columns how many percentage of the ensemble could learn the problem within the simulation time of $T=10^7$ time steps. One can see that also in this category learning rule 4.b) is better then 4.a) because the convergence percentage is alway greater in any parameter region of $p_\mathrm{rw}$. However, for $p_\mathrm{rw}\le 0.3$ even learning with rule 4.b) is not perfect.

\begin{table}[h!]
\caption{\label{list_ens1}Ensemble sizes used for the corresponding simulations and the percentage of networks which learned the mapping after $T=10^7$ time steps correctly (ensemble size/percentage). LR means learning rule. The action-selection-mechanism which was used in these simulations was ASM I.}
\begin{ruledtabular}
\begin{tabular}{c|cccc} \hline
 & LR 4.b) & LR 4.b) & LR 4.a) & LR 4.a) \\
$p_\mathrm{rw}$&M=3&M=5&M=3&M=5\\ \hline
0.1&1000/52.7&100/0&500/23.6&-    \\
0.2&1000/94.7&100/3.5&500/75.0&-   \\
0.3&1000/99.9&100/12&500/92.6&-     \\
0.4&1000/100&100/30.9&500/97.8&-    \\
0.5&1000/100&500/48&500/99.4&-     \\
0.6&1000/100&500/63.2&500/99.8&-    \\
0.7&1000/100&500/80&500/100&-    \\
0.8&1000/100&500/87.1&500/100&100/0.0    \\
0.9&1000/100&500/94.7&500/100&100/3.7    \\
1.0&1000/100&500/96.2&500/100&100/4.4     \\   \hline
\end{tabular}
\end{ruledtabular}
\end{table}

\begin{figure}[h!]
\begin{minipage}[c]{0.4\textwidth}
\epsfig{file=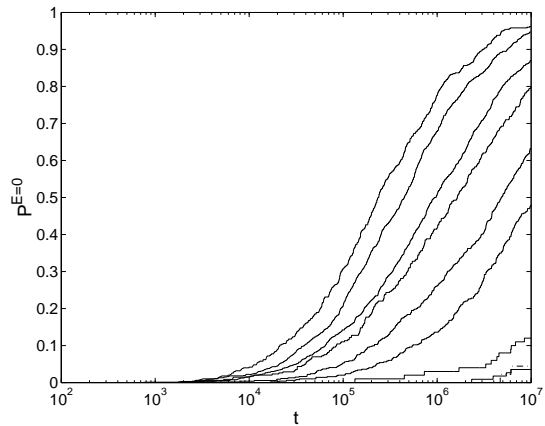, width=1.0\textwidth}
\end{minipage}
\caption{\label{asm1_m5_p}Distribution function $P^{E=0}(t)$ for $M=5$ and learning rule 4.a) (dash-dot line) and 4.b) (full line) which were obtained for ASM I. The curves for learning rule 4.b) are parametriesed from above to below from $p_\mathrm{rw}=1.0$ to $p_\mathrm{rw}=0.1$ and for learning rule 4.a) from $p_\mathrm{rw}=1.0$ to $p_\mathrm{rw}=0.9$.}
\end{figure}

{\bf{$M=5$ patterns:}}\\
In figure \ref{asm1_m5_p} the corresponding results for $M=5$ patterns are shown. Here the effects mentioned above are further increased by increasing the number of patterns to be learned. This results in an almost complete break down for learning rule 4.a) which is now only able to learn the problem for $p_\mathrm{rw}=\{0.9,1.0\}$ for a few networks. Learning rule 4.b) works much better also in this case. However, learning within $T=10^7$ time steps is always incomplete.

\subsection{Action-selection-mechanism II.}

The results for action-selection-mechanism II. and $M=3$ respectively $M=5$ patterns are summarized in table \ref{list_ens2}. 
\begin{table}[h!]
\caption{\label{list_ens2}Ensemble sizes used for the corresponding simulations and the percentage of networks which learned the mapping after $T=10^7$ time steps correctly (ensemble size/percentage). LR means learning rule. The action-selection-mechanism which was used in these simulations was ASM II.}
\begin{ruledtabular}
\begin{tabular}{c|cccc}\hline
& LR 4.b) & LR 4.b) & LR 4.a) & LR 4.a) \\
$p_\mathrm{rw}$&M=3&M=5&M=3&M=5\\ \hline
0.1&1000/57.4&500/1.0&500/31.2&-\\
0.2&1000/97&500/7.2&500/74.4&-\\
0.3&1000/99.9&500/30.5&500/91.4&-\\
0.4&1000/100&500/53.3&500/97.6&-\\
0.5&1000/100&500/76.6&500/99.8&-\\
0.6&1000/100&500/85.6&500/99.6&100/0.0\\
0.7&1000/100&500/92.4&500/99.8&100/1.0\\
0.8&1000/100&500/97.4&500/100&100/4.0\\
0.9&1000/100&500/97.4&500/100&100/3.0\\
1.0&1000/100&500/99.0&500/100&100/6.0\\  \hline
\end{tabular}
\end{ruledtabular}
\end{table}
One recognizes by comparison with table \ref{list_ens1} that the overall results are confirmed. Learning rule 4.b) obtains always significantly better results than 4.a). Moreover, a direct comparison between the learning rules for ASM I. and II. reveals that learning rule 4.a) seems to be unaffected by the action-selection mechanism whereas learning rule 4.b) is clearly influenced.

\subsection{Comparison of ASM I. and ASM II.}

\begin{figure}[h!]
\begin{minipage}[c]{0.45\textwidth}
\epsfig{file=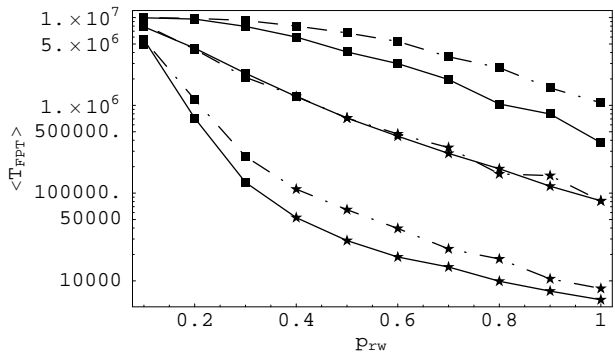, width=1.0\textwidth}
\end{minipage}
\caption{\label{meanfpt_rs}Mean first-passage time $<T_\mathrm{FPT}>$ in dependence of the rewiring parameter $p_\mathrm{rw}$. Full lines correspond to simulations with ASM II. dashed-dotted lines to ASM I. The two upper (lower) curves are obtained by learning rule 4.b) and  $M=5$ ($M=3$) patterns. The two middle curves by learning rule 4a) and $M=3$ patterns. ``$*$'' indicates that $100\%$ of the ensemble converged within the simulation time of $10^7$ time steps whereas ``$\Box$'' indicates that this is not the case. This implies that the obtained results for $<T_\mathrm{FPT}>$ are only estimations.}
\end{figure}

To quantify the dependence of the learning behavior of the action-selection mechanism we calculate the mean first-passage time $<T_\mathrm{FPT}>$ from the simulation results obtained so far. Figure \ref{meanfpt_rs} compares the results for learning rule 4.a) and 4.b) in dependence of the rewiring parameter $p_{rw}$ and the patterns to be learned. 
One can clearly see that the mean first-passage time for learning rule 4.b) (upper (lower) two curves correspond to $M=5$ ($M=3$)) is significantly reduced for ASM II. (full lines) whereas the results for learning rule 4.a) are not affected (middle curves correspond to $M=3$).

This can be explained by the different structure of both learning rules. Learning rule 4.a) possesses no memory with respect to the outcomings of past results but only a tagging mechanism for the neurons which were involved in the last signal processing step. Hence it can not detect the differences of the two action-selection mechanisms because they differ only in the order of the presented patterns but not in the overall presentation statistics. This follows from the fact that learning the last pattern takes about $90\%$ of the first-passage time. Learning rule 4.b) is due to the neuron counters $c_i$ different in this point. The neuron counters are a memory for the outcomings of the past results and thus can detect the slight difference in the two action-selection mechanisms.

We think that this result is worth to be discussed in detail because it reveals some deep characteristics of animals which is normally neglected in investigations of neural networks. The consequences of the results obtained above are not only that the learning rule of a neural network effects on the neural activity by synaptic changes and hence on the behavior of an animal which is common sense, but also that the reverse holds. That means the actions of an animal influence the learning rule of its neural network. This is caused by the stimuli generated by the animal's actions which are represented in the examples above as patterns which lead to a modulation of the neural activity in the network and hence to a modulation of the learning rule due to memory effects by the neuron counters. This seems to be plausible because we do not choose our actions randomly but we choose them to learn something as fast as possible to survive. Moreover, it is not only plausible but also efficient to us the action-selection mechanism as source of information which is shown in figure \ref{meanfpt_rs}.

Hence our investigations lead not only to a bottom-up communication but also to a top-down communication between different system levels. In this respect our learning rule with neuron counters is different to all other Hebb-like learning rules which has been proposed as extentions to the classical Hebbian rule \cite{h1949} which lack the ability of a memory because they can not be affected by action-selection mechanisms which differ not in the presentation statistics but only in the presentation order.


\section{Conclusions}\label{conclusions}

In this article we investigated the properties of our recently proposed stochastic Hebb-like learning rule for neural networks. We demonstrated by extensive numerical simulations that the problem of timing can be learned in different topologies of a neural network generated by the algorithm of Watts and Strogatz \cite{WattsStro_1998}. A comparison with the learning rule of Chialvo and Bak \cite{cb1999,bc2001} gave not only always significantly better results but revealed that our stochastic Hebb-like learning rule can discriminate between different action-selection mechanisms with the same presentation statistics but different presentation order. This difference forms a source of information and can positively effect the learning behavior due to the bidirectional communication between different system levels. This effect was only recognized because we did not want to model the brain of an animal but its action-perception-cycle schematically depicted in table \ref{gen_sys} where the brain is only one part of the entire system.

In summary our stochastic Hebb-like learning rule is not only universal applicable in feedforward multilayer networks \cite{vhebb1_2003} but also in a class of recurrent networks generated by \cite{WattsStro_1998} as demonstrated in this article. Together with its biological interpretation as qualitative form of heterosynaptic plasticity \cite{vphd_2003,vhebb1_2003} and its sensitivity to the presentation order of the patterns to be learned we belief that our learning rule unites some crucial ingredients on the way of our understanding of the action-perception-cycle and hence of the brain. We belief that only such an integrated ansatz can explain the functional working method of the entire system because its parts are coupled in a nonlinear or stochastic way.


\begin{acknowledgements}
We would like to thank Rolf D. Henkel and Klaus Pawelzik for fruitful discussions and Tom Bielefeld for carefully reading the manuscript.
\end{acknowledgements}



\end{document}